# Possible, Impossible, and Expected Diameters and Production Rates of Droplets in Aerosols and Sprays

Maksim Mezhericher* and Howard A. Stone

Department of Mechanical and Aerospace Engineering, Princeton University, Princeton, USA

*Corresponding author email: maksymm@princeton.edu

## Abstract

Liquid atomization processes generating sprays and aerosols of droplets are used in many delivery and coating systems involving pure solvents, solutions, and suspensions. In our recent experimental work, we introduced a novel liquid atomization process generating micro-sprays and aerosols of submicron-diameter droplets for pure solvents, solutions, and suspensions: gas jets disintegrate thin liquid films that are formed as bubbles approach a liquid surface. Here we develop a theoretical description of droplet sizes and flow rates, using the first principles of conservation of mass and energy, and accounting for the ratios of specific energies and the ratios of specific energy rates provided by the atomizing gas and dissipated by the atomized liquid. We introduce atomization diagrams as a graphical tool to determine possible, impossible, and expected droplet diameters and specific flow rates in aerosols and sprays produced under various conditions. We find a reasonable agreement between the theory and experiments for five different liquids converted into aerosols of submicron-diameter droplets by an atomization process where gas jets disintegrate thin liquid films that are formed as bubbles approach a liquid surface, and also for five traditional pressure nozzles that produce sprays of droplets of hundreds of microns in diameter. Our study explored the overall range of Ohnesorge number between 0.01-100, and the analysis and atomization diagrams contribute to understanding of liquid atomization and can serve as a theoretical framework for comparing different liquid atomization techniques.

Disintegration of liquids into drops plays a central role in many natural processes as well as in various technological, industrial, and medical applications. Wind and bubble-driven droplet generation over sea surfaces [1,2], splashing of raindrops [3–6], defense mechanisms of some insects [7], spray painting [8], coating of surfaces and particles [9,10], agricultural treatment of plants [11,12], fuel injection during combustion [13,14], atomic spectroscopy [15,16], instant coffee [17,18] and milk powder [19] manufacturing, nasal [20,21] and pulmonary [22,23] drug delivery, blood spatter [24], dental procedures [25], spreading of airborne pathogens while speaking [26–28], singing [26], sneezing [29,30] and coughing [30,31], and other processes involve production of aerosols and sprays by atomization of liquids. The range of applications is clearly varied and enormous.

In many applications the quality, controllability, stability, reproducibility, and scalability of the atomization process is crucial, and scientists seek to understand the physics of liquid disintegration and dispersion and develop methods and models to improve droplet generating devices. In previous articles [10,32–34] we demonstrated a novel liquid atomization process (Fig. 1a). The process utilizes gas jets to disintegrate bubble-formed thin liquid films, which enables production of polydisperse aerosols of submicron droplets [35] for pure solvents, solutions, and suspensions with wide ranges of viscosity and surface tension.

For most droplet production process, an important but unanswered question is how to predict the droplet diameters and flow rates that will be produced for different liquids during atomization. Another question is how to compare between different atomization processes. The available review literature on liquid atomization and spray and aerosol formation processes [36–39] demonstrates the absence of an appropriate theoretical framework to answer these questions because of complexity of the fluid dynamics involved in the formation of polydisperse sprays and aerosols with time-dependent droplet size distributions [40–46]. Currently the description of liquid atomization and corresponding devices (atomizers, nozzles and nebulizers) is mostly empirical and semi-empirical [37,47,48], or based on numerical modeling utilizing approaches (e.g., Eulerian/Lagrangian, Reynolds-averaged Navier-Stokes, direct numerical simulations, large-eddy simulations) that are filled with questionable assumptions and require high computational costs that limit their applicability [49–51]. Therefore, there is a need to develop a theoretical description providing determination of possible, impossible, and expected droplet diameters and flow rates produced by a liquid atomization process, and enabling comparison between different liquid atomization techniques.

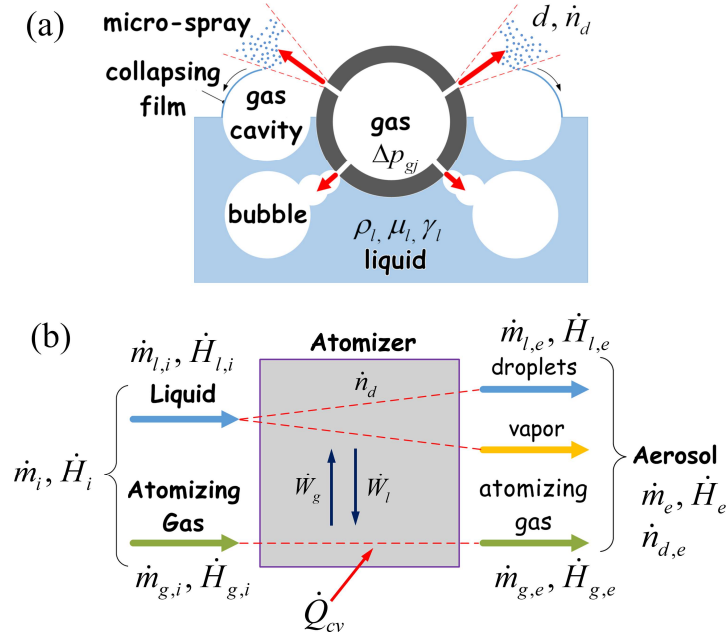

FIG. 1. (a) Liquid atomization process generating submicron droplets by gas jet disintegration of thin liquid films formed as bubbles on a liquid surface [10,32–34]. (b) Control volume for thermodynamic analysis of the liquid atomization process. In panel (a), compressed gas is supplied into a perforated tube (the tube cross section is shown) partially submerged into a liquid in such way that the underwater part produces gas bubbles, which rise to the liquid surface. Gas jets, which are produced in the upper part of the perforated tube, disintegrate bubbles over the liquid surface [32,33]. Here $d$ is the droplet diameter, the pairs $\dot{m}_{l,i}, \dot{H}_{l,i}$, $\dot{m}_{g,i}, \dot{H}_{g,i}$ and $\dot{m}_i, \dot{H}_i$ are, respectively, the mass flow rate and rate of change of enthalpy for liquid, gas, and total flow at the inlet; the pairs $\dot{m}_{l,e}, \dot{H}_{l,e}$, $\dot{m}_{g,e}, \dot{H}_{g,e}$ and $\dot{m}_e, \dot{H}_e$ are, respectively, the analogous quantities at the exit; $\dot{n}_d, \dot{n}_{d,e}$ are, respectively, the rate of generated droplets inside and at the exit; $\dot{Q}_{cv}$ is rate of heat supply; $\dot{W}_g$ and $\dot{W}_l$ are rates of work, respectively, done by the gas phase and dissipated by the liquid phase inside the control volume.

The atomization process is analyzed by using mass conservation and the first law of thermodynamics for a control volume with an imaginary boundary drawn over an atomizing device (Fig. 1b). Assuming steady-state flow for a continuous adiabatic process, neglecting changes in potential energy, evaporation and the change in liquid kinetic energy, assuming ideal expansion of the gas jet, and neglecting drag, we obtain that mass flow rates at the inlet and outlet are equal and the net work done by the control volume is zero. Therefore, inside the control volume the rate of work supplied by the gas phase equals to the rate of work done by the liquid phase, $\dot{W}_g = -\dot{W}_l$. In turn, the latter is determined by the sum of work rates done by capillary, $\dot{W}_c$, and viscous, $\dot{W}_v$, forces of the liquid, $\dot{W}_l = -\left(\dot{W}_c + \dot{W}_v\right)$. Thus, we obtain that $\dot{W}_g = \dot{W}_c + \dot{W}_v$, and dividing by the liquid (droplet) mass flow rate, $\dot{m}_l$, we get:

$$w_{\Delta p} = w_c + w_v ,\tag{1}$$

where $w_{\Delta p}$ is the specific work (per unit mass) done by the gas phase on the liquid phase because of the supplied differential pressure $\Delta p_{gj}$ to produce gas jets (Fig. 1a), and $w_c$ and $w_v$ are, respectively, the specific work done by capillary and viscous forces in the atomized liquid.

In previous research [32], we established two governing dimensionless groups: Ohnesorge number, $\text{Oh}_d = \mu_l / \sqrt{\rho_l \gamma_l \ell}$, and $\text{N}_d = \Delta p_{gj} \ell / \gamma_l$. Here $\rho_l$, $\gamma_l$, and $\mu_l$ are, respectively, liquid density, surface tension, and dynamic viscosity, and $\ell$ is a characteristic length. These two dimensionless numbers are central to determination of droplet diameters produced in a liquid atomization process. Using the timescales of the liquid atomization, including the timescale of energy supplied by the gas jets $\tau_{\Delta p} \sim \left(\rho_l \ell^2 / \Delta p_{gj}\right)^{1/2}$, the Rayleigh capillary breakup time $\tau_c \sim \left(\rho_l \ell^3 / \gamma_l\right)^{1/2}$ and the timescale of viscous dissipation $\tau_v \sim \rho_l \ell^2 / \mu_l$ in the liquid, the dimensionless groups can be expressed as timescale ratios, $\text{Oh}_d^{-2} \sim \dfrac{\tau_v^2}{\tau_c^2}$ and $\text{N}_d \sim \dfrac{\tau_c^2}{\tau_{\Delta p}^2}$. Alternatively, the energy scales and timescales are connected, i.e., the scales of the energy associated with the gas pressure $e_{\Delta p} = \ell^2 / \tau_{\Delta p}^2$, the energy stored or released by surface effects $e_c = \ell^2 / \tau_c^2$, and the energy dissipated by liquid viscosity $e_v = \ell^2 / \tau_v^2$, so that $\text{Oh}_d^{-2} \sim \dfrac{e_c}{e_v}$ and $\text{N}_d \sim \dfrac{e_{\Delta p}}{e_c}$. Assuming that energy scales are proportional to the respective specific works, then $e_c \sim w_c$, $e_v \sim w_v$, and $e_{\Delta p} \sim w_{\Delta p}$. Correspondingly, we conclude that physical meaning of the dimensionless numbers are ratios of specific work, and $\mathrm{Oh}_d^{-2} = k_1 \dfrac{w_c}{w_v}$ and $\mathrm{N}_d = k_2 \dfrac{w_{\Delta p}}{w_c}$, where $k_1$ and $k_2$ are coefficients of proportionality. Substituting these expressions into Eq. (1), we establish a relationship between the dimensionless numbers:

$$\mathrm{N}_d = k_2 + k_1 k_2 \mathrm{Oh}_d^2 \qquad (2)$$

Equation (2) relates the specific works performed by the gas jets and dissipated by the atomized liquid, and enables constructing a diagram for determination of droplet diameters on the plane $\left(\mathrm{Oh}_d^{-2}, \mathrm{N}_d\right)$. Though the proportionality coefficients are unknown, there are either empirical or theoretical ways to determine their values. Our theoretical study, reported elsewhere [52], which involves the development and solution of a stochastic differential equation for the droplet size distribution function, suggests that in general $O(k_1) = O(k_2) = 1$, and provide a theoretical framework to determine the constant $k_2$. Here we do not apply that complicated stochastic method, yet we investigate different values of proportionality coefficients for the range $k_2 = 0.1 - 10$ while keeping $k_1 = 1$, and compare the predicted droplet diameters with broad experimental data.

We assume that the characteristic length is equal to a droplet diameter, $\ell = d$, and construct an atomization diagram for water, when the droplet diameter, $d$, and gas jet differential pressure, $\Delta p_{gi}$, are varied, see Fig. 2a. The central line is given by the relationship $\mathrm{N}_d^* = 1 + \mathrm{Oh}_d^2$, which is obtained from (2) by setting $k_1 = k_2 = 1$ as a first approximation. This line provides the expectation of droplet diameters at different atomizing gas pressures $\Delta p_{gi}$. In a simple case, the expected diameter for a polydisperse aerosol or spray, $\langle d \rangle$, is the "count" mean droplet diameter, $\overline{d}_{1,0}$, of the respective number-weighted droplet size distribution, i.e., $\langle d \rangle = \overline{d}_{1,0}$ [36,38]. For water and common aqueous solutions, $\mathrm{Oh}_d^2 \ll 1$, so the central line in the atomization diagram is given by the limit $\mathrm{N}_d^{**} = 1$. Recalling that $\mathrm{N}_d \sim \dfrac{e_{\Delta p}}{e_c}$, we outline the atomization region of possible droplet diameters by considering $e_{\Delta p} \approx e_c$, i.e., $O\left(\mathrm{N}_d\right)=1$, and determine two boundaries of the atomization region: the lower boundary, below which the supplied energy is insufficient to perform atomization, $e_{\Delta p} \ll e_c$; and the upper boundary, above which there is an excess of atomization energy, $e_{\Delta p} \gg e_c$, meaning that a liquid particle there will be disintegrated into smaller fragments. These two conditions can be satisfied by considering an order of magnitude difference from the central line, so we take $\mathrm{N}_d = 0.1\mathrm{N}_d^{**}$ and $\mathrm{N}_d = 10\mathrm{N}_d^*$ as, respectively, the lower and the upper boundaries of the atomization region (Fig. 2a). It is worth noting that those two boundaries encompass a wide scope of the coefficients of proportionality, i.e., $\mathrm{N}_d = 0.1\mathrm{N}_d^{**}$ is equivalent to setting $k_2 = 0.1$, and $\mathrm{N}_d = 10\mathrm{N}_d^*$ is equivalent to setting $k_2 = 10$ in Eq. (2).

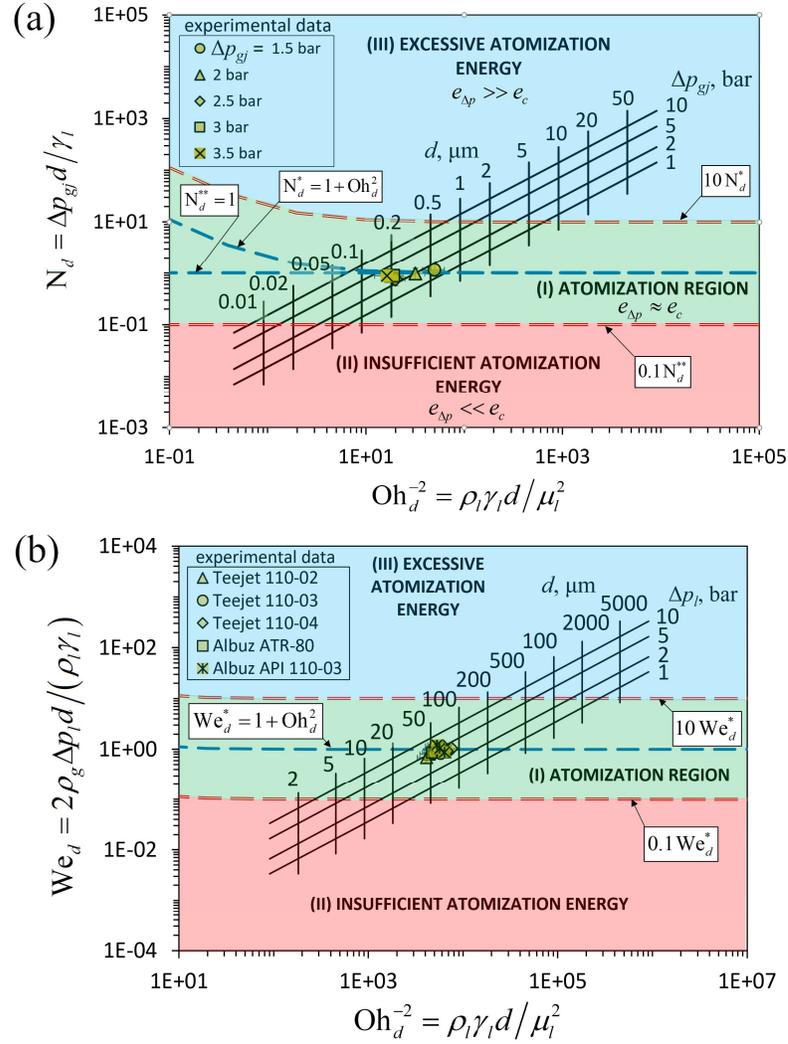

FIG. 2. Atomization diagrams for the diameters of water droplets generated in two different liquid atomization processes. (a) Aerosols of water droplets generated by our own process (Fig. 1a); the experimental data (symbols) were obtained for different atomizing gas pressures in our previous study [33]. (b) Sprays of water droplets produced by conventional pressure nozzles; labels correspond to experimental data (symbols) for various nozzle models reported in [43]. Error bars denote measurement uncertainties as we evaluated here (experimental errors were not provided in [43]). Region (I) encompasses the area of the expected droplet diameters, and droplets in the regions (II) and (III) have either too small or too large diameters with respect to the atomization energy that is required to overcome the work of capillary and viscous forces in the liquid.

Using the data on the count mean diameter, $\bar{d}_{1,0}$, for water droplets measured by Malvern Spraytec device [33], we plot the experimental points on the atomization diagram given in Fig. 2a. It can be observed that those points either fall on or lie very close to the expectancy line, $N_d^{**} = 1$, and hence there is a good agreement between the theoretically expected and measured count mean droplet diameters using $O(1)$ proportionality constants.

The theoretical procedure for constructing atomization diagrams for droplet diameters can also be applied to other liquid atomization processes, for example pressure nozzles as given in Fig. 2b. The diagram was constructed on the plane $\left( \text{Oh}_d^{-2}, \text{We}_d \right)$, where $\text{We}_d = \rho_g u_{rel}^2 d / \gamma_l = 2 \rho_g \Delta p_l d / \left( \rho_l \gamma_l \right)$ is the Weber number for an ideal nozzle discharge into a still ambient gas based on the droplet diameter, $d$, and relative velocity $u_{rel} = \sqrt{2 \Delta p_l / \rho_l}$ between the liquid jet exiting the nozzle and the ambient gas, and $\Delta p_l$ is nozzle pressure. Using the steps we took to develop Eq. (2), we obtain for pressure nozzles:

$$\text{We}_d = k_4 + k_3 k_4 \text{Oh}_d^2 \qquad (3)$$

As above, to construct the atomization diagram in Fig. 2b, we consider that the proportionality coefficients are order of one, $O\left( k_3 \right) = O\left( k_4 \right) = 1$, and obtain the expectancy line $\text{We}_d^* = 1 + \text{Oh}_d^2$ by setting $k_3 = k_4 = 1$ as a first approximation. Overall, we investigate different values of proportionality coefficients by setting $k_4 = 0.1, 1, 10$ while keeping $k_3 = 1$ and comparing the predictions with experimental data. The lower boundary of the atomization region, below which there is insufficient atomization energy to overcome the energy associated with capillary forces, $e_{\Delta p} \ll e_c$, and the upper boundary of the atomization region above which there is an excess of atomization energy, $e_{\Delta p} \gg e_c$, are determined by assuming tenfold ratios of the expected Weber number, respectively $0.1 \text{We}_d^*$ and $10 \text{We}_d^*$. It is worth noting that those two boundaries encompass a wide scope of the coefficients of proportionality, i.e., $0.1 \text{We}_d^*$ is equivalent to setting $k_4 = 0.1$, and $10 \text{We}_d^*$ is equivalent to setting $k_4 = 10$ in Eq. (3). Also, for water and common aqueous solutions with $\text{Oh}_d^2 \ll 1$, the expectacy line expression from Eq. (3) is reduced to $\text{We}_d^{**} = 1$, and thus the upper boundary of the atomization region is given by the line $10\mathrm{We}_d^{**} = 10$. Such a value of the upper boundary is justified by the well known theoretical [53,54] and experimental observations [36,37,55] in liquid atomization that droplets with $\mathrm{Oh}_d \ll 1$ and $\mathrm{We}_d > 10$ are unstable and undergo so-called "secondary atomization" caused by aerodynamic drag breakup mechanisms [48,56]. On the other hand, droplets with $\mathrm{Oh}_d^2 \ll 1$ and Weber number below the lower boundary line of atomization region, $0.1\mathrm{We}_d^{**} = 0.1$, may undergo only small oscillatory shape deformation without breakup [48].

Comparing Eqs. (2) and (3), we conclude that the Weber number for sprays plays a similar role to the dimensionless group $\mathrm{N}_d$ established before for aerosol production. Using the measured mass median droplet diameters of water sprays, $\bar{d}_{50}$, reported in the study [43] for different pressure nozzles, employing the Hatch-Choate [38,57] relationship $\bar{d}_{1,0} = \bar{d}_{50} \exp\left[ -2.5\left(\ln \sigma_g\right)^2 \right]$ for a lognormal droplet size distribution, and taking the geometric standard deviation $\sigma_g = 1.7$ (based on Fig. 17 in [43]), we plot the experimental data on Fig. 2b. All the experimental points have $\mathrm{Oh}_d^2 \ll 1$ and either fall on or lie very close to the expectancy line, $\mathrm{We}_d^{**} = 1$. Again, we conclude that there is good agreement between the theoretically expected and measured count mean droplet diameters with $O(1)$ proportionality constants.

Atomization diagrams can also be constructed for droplet flow rates. For this purpose, we take the time derivative of Eq. (1), employ the dimensionless analysis described in our previous article [32] to establish the two dimensionless numbers governing droplet flow rate for our liquid atomization process, $\mathrm{N}_{l,cv} = \rho_l^{5/2} \gamma_l^{3/2} d^{9/2} \xi / \mu_l^3$ and $\mathrm{N}_l = \Delta p_{gj}^{3/2} / \left( \rho_l \gamma_l^{3/2} d^{3/2} \xi \right)$, and use scaling analysis to find:

$$\mathrm{N}_l = k_6 + \frac{k_5 k_6}{\mathrm{N}_{l,cv}} \tag{4}$$

Here $\xi = \dot{n}_{d,e} / \dot{m}_{d,e}$ is the specific flow rate of droplets for the expected droplet diameters $\mathrm{N}_d^{**} = \Delta p_{gj} \langle d \rangle / \gamma_l = 1$, and $\dot{n}_{d,e}$ and $\dot{m}_{d,e}$ are, respectively, the number and the mass flow rates of droplets at the outlet of the atomizer (see Fig. 1b). The established Eq. (4) relates the rates of specific work provided by the gas jets and dissipated by the atomized liquid in nondimensional form, and determines the specific flow rates of produced droplets. This model was applied to assemble the atomization diagram shown in Fig. 3a by varying $k_6 = 0.1 - 10$ and keeping $k_5 k_6 = 1$. Using the experimental data [33], we observe good agreement between the theoretically expected and measured specific droplet flow rates; the experimental points either fall on or lie very close to the expectancy line, $N_l^{**} = \pi/6$ (more details and diagrams for additional liquids are in the Supplementary Material).

The procedure for constructing an atomization diagram for droplet flow rates was also applied to pressure nozzles and the results for water spray droplets are shown in Fig. 3b. The diagram was developed for the plane $\left( N_{l,cv}, N_l \right)$, with the dimensionless numbers $N_{l,cv} = \rho_l^{5/2} \gamma_l^{3/2} d^{9/2} \xi / \mu_l^3$ and $N_l = \left( 2 \rho_g \Delta p_l \right)^{3/2} \Big/ \left( \rho_l^{5/2} \gamma_l^{3/2} d^{3/2} \xi \right)$ established by employing the dimensional analysis for a pressure nozzle. Also, Fig. 3b demonstrates the experimental data based on the diameters and flow rates of water droplets for different models of pressure nozzles reported in [43]. All the experimental points are located in the atomization region corresponding to the expected droplet flow rates, and are concentrated either on or near the expectancy line, $N_l^{**} = \pi/6$.

To extend our study, we have constructed atomization diagrams for droplet diameters of other liquids, including gasoline, diesel and aqueous solutions of sodium alginate and sodium benzoate, exploring the overall range of Ohnesorge number between $0.01–100$ ($Oh_d^{-2} = 10^{-4} - 10^2$, see details in the Supplementary Material). Figure 4a demonstrates summarizing atomization diagram for droplet diameters of all the studied liquids, and the summarizing diagram for specific flow rates is given in the Supplementary Material. The expected and measured mean droplet diameters for various liquids from different atomization diagrams are shown in Fig. 4b,c for both our liquid atomization process and for spraying by pressure nozzles reported in the study [43], and the summary diagram for specific flow rates is given in the Supplementary Material.

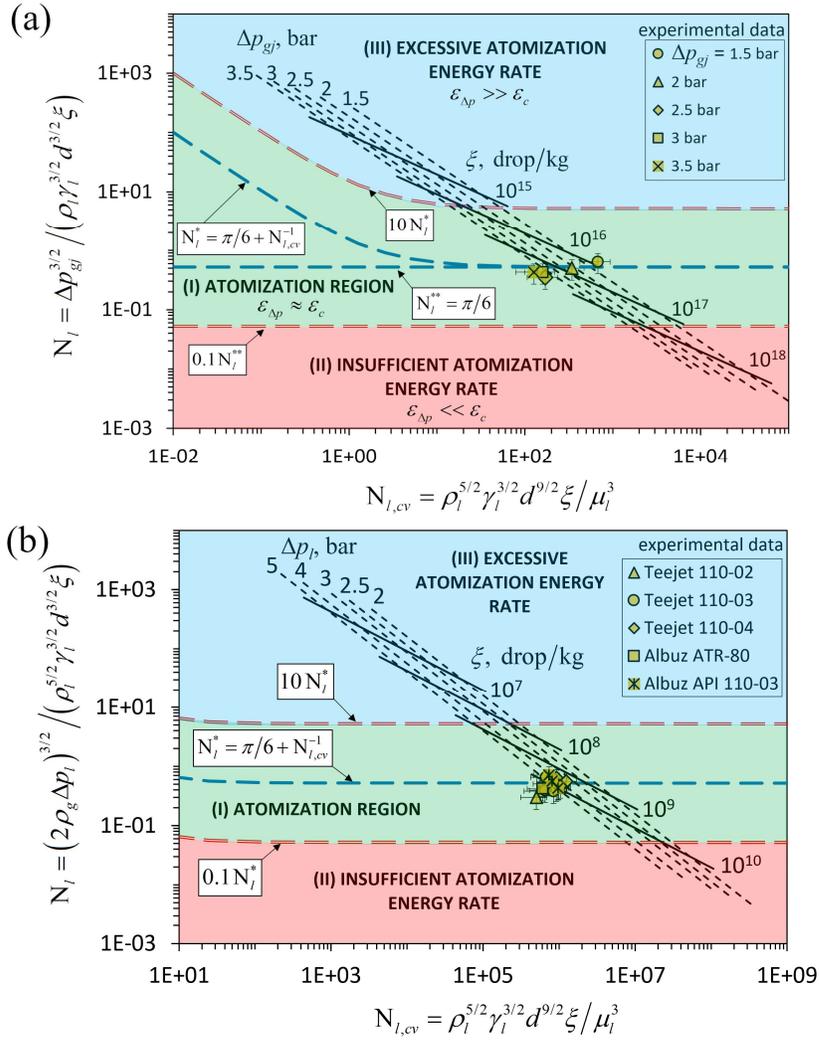

FIG. 3. Atomization diagrams for specific flow rates of water droplets generated in two different liquid atomization processes. (a) Aerosols of water droplets generated by our atomization process (Fig. 1a); the experimental data (symbols) were obtained for different atomizing gas pressures [33]. (b) Sprays of water droplets produced by pressure nozzles; labels correspond to experimental points (symbols) reported for various nozzle models in [43]. The experimental points are based on the measured count mean droplet diameters, $\bar{d}_{1,0}$, and error bars denote the evaluated measurement uncertainties. Region (I) encompasses the area of the expected specific droplet flow rates, and droplets in the regions (II) and (III) have either too large or too small specific droplet flow rates with respect to the invested atomization energy rate required to overcome the work rates of capillary and viscous forces in the liquid.

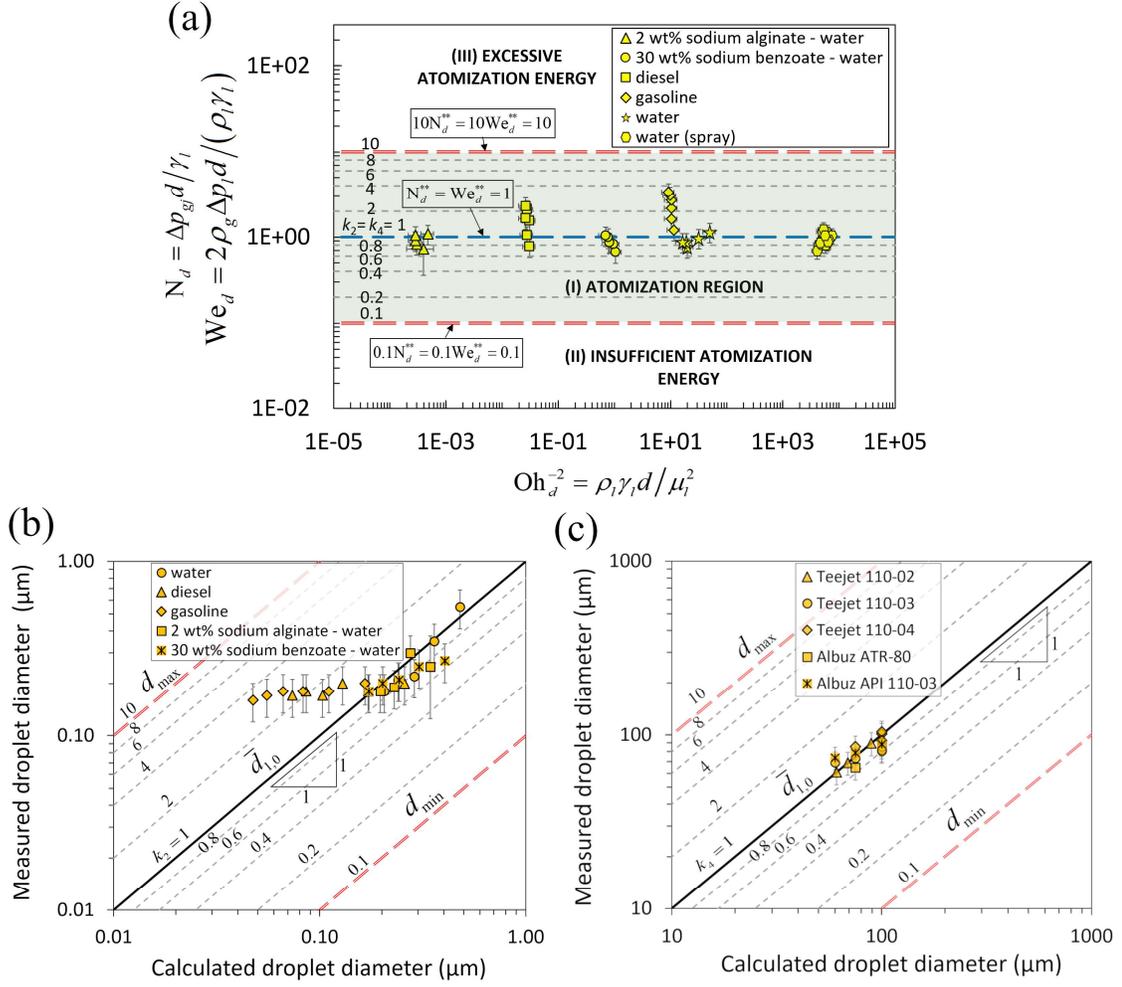

FIG. 4. (a) Atomization diagram summarizing droplet diameters produced by the two studied atomization processes, aerosol generation shown in Fig. 1a [32,33] and spraying from pressure nozzles [43]. (b) Comparison between theoretical predictions and experimental data [32,33] for count mean droplet diameters, $\bar{d}_{1,0}$, of various liquids subjected to the atomization process shown in Fig. 1a. (c) Comparison between theoretical predictions and experimental data [43] for count mean droplet diameters, $\bar{d}_{1,0}$. The lines $\bar{d}_{1,0}$ correspond to the lines of the expected dimensionless numbers $N_d^*$ and $We_d^*$ in the atomization diagrams of Figs. 2 and 3 (obtained by setting proportionality coefficients $k_1 = k_2 = k_3 = k_4 = 1$ in Eqs. (2) and (3)). The double-dashed lines denote theoretical lower and upper boundaries of possible droplet diameters, $d_{min}$ and $d_{max}$, for the produced polydisperse aerosols and sprays, and correspond to the boundary lines of the atomization regions in Figs. 2 and 3 (obtained by setting each of $k_2$ and $k_4$ to 0.1 and 10 while keeping $k_1 = k_3 = 1$ in Eqs. (2) and (3)). Error bars indicate the evaluated measurement uncertainties.

In general, the results in Fig. 4 demonstrate an agreement between the theoretically expected and measured mean droplet diameters and specific flow rates for both droplet aerosols and sprays. Most of the experimental points, within the measurement uncertainties, are concentrated either on or near the respective expectancy lines (central line $k_2 = k_4 = 1$ in Fig. 4a and line of $\overline{d}_{1,0}$ in Figs. 4b,c), within the range of the expected droplet diameters, respectively, in between the boundaries $d_{\min}$ and $d_{\max}$. However, when the calculated mean droplet diameters are smaller than 0.1 μm, the results in Figs. 4b (and respectively in Fig. 4a) show deviations of some experimental points from the theoretically expected values for droplets of gasoline and diesel. Because these fuels had the smallest values of surface tension among all the studied liquids (17 and 26 mN/m, see Table S1 in Supplementary Material), their atomization produced aerosols with the smallest droplet diameters [32]. The measurement device (Malvern Spraytec) was unable to resolve droplets with diameters smaller than 0.1 μm [32,58] due to the optical limit of the utilized Mie scattering laser technique [59,60]. As a result, the measured droplet size distributions did not adequately represent the fractions of small droplets by disregarding droplets with diameters <0.1 μm in fine polydisperse aerosols or sprays. Hence, there is a systematic bias in the experimental mean droplet diameters, which shifted the data towards higher values, and this bias grows with increase of the fraction of sub-0.1 μm droplets in the droplet size distributions, as observed from Fig. 4b. Therefore, the experimental data encompassing Ohnesorge numbers 0.01–100 suggest that the proportionality coefficients $k_1,...,k_4$ are close to unity, and thus setting $k_1 = k_2 = k_3 = k_4 = 1$ is a reasonable choice at a first approximation to find the expected, possible and impossible droplet diameters of a steady adiabatic liquid atomization process with negligible droplet evaporation, without a need to solve detailed differential equations of the process.

The theoretical framework and atomization diagrams can serve as a tool for comparison between liquid atomization techniques. As an example, we compare droplets of water produced at the same differential pressure applied in the two considered atomization systems. Setting the differential nozzle pressure equal to the atomizing gas pressure drop, $\Delta p_l = \Delta p_{gl}$, and equating the governing dimensionless numbers in both systems, $N_d = We_d$, while the Ohnesorge number is the same, thus we find that the expected droplet diameters for a conventional pressure nozzle are ~400 times bigger than those for our liquid atomization device (cf. Fig. 4b and Fig. 4c). Also, the expected number of droplets generated from 1 kg of water by pressure nozzles is several orders of magnitude smaller than for our liquid atomization process producing aerosols of submicron-diameter droplets, because $\xi \sim 1/d^3$ (see Supplementary Material).

The developed theoretical approach assumes that energy supplied to an atomization process is completely spent overcoming surface tension and viscous forces, which is needed to disintegrate bulk liquid into droplets. For many atomization processes under normal conditions additional dissipation effects can be negligible [36,43,44], but for some atomization processes or under certain (usually extreme) conditions these other dissipation mechanisms can impact the droplet diameters and flow rates [44]. In this sense, our theoretical model describes an idealized liquid atomization process, and predicts possible, impossible, and expected droplet diameters and specific flow rates accounting only for surface tension and viscous effects, without considering additional irreversible energy dissipation phenomena.

## Acknowledgments

The research reported in this publication was supported by the Foundation for Health Advancement and New Jersey Health Foundation via the Allergan Foundation Innovation Grant Program, Grant # ALL 01-21, and the National Center for Advancing Translational Sciences (NCATS), a component of the National Institute of Health (NIH) under award number UL1TR003017. The content is solely the responsibility of the authors and does not represent the official views of the National Institutes of Health.

# SUPPLEMENTARY MATERIAL

# Possible, Impossible, and Expected Droplet Diameters and Flow Rates in Aerosols and Sprays

Maksim Mezhericher* and Howard A. Stone

Department of Mechanical and Aerospace Engineering, Princeton University, Princeton, USA

*Corresponding author, email: maksymm@princeton.edu

**Table S1**. Properties of liquids used in experiments (adapted from [1–3])

| Liquid | Density, kg/m$^3$ | Surface tension*, mN/m | Viscosity, mPa·s |
|---|---|---|---|
| diesel | 814±19 | 25.8±1.1 | 11.66±0.06 |
| gasoline (95 RON) | 734±19 | 16.6±0.9 | 0.46±0.01 |
| 2 wt.% sodium alginate in water | 1004±15 | 68.9±2.4 | 208.60±1.04 |
| 30 wt.% sodium benzoate in water | 1125±17 | 60.7±1.8 | 4.19±0.09 |
| water | 998 | 72.86 | 1.00 |

*The values of surface tension were obtained in air at room temperature.

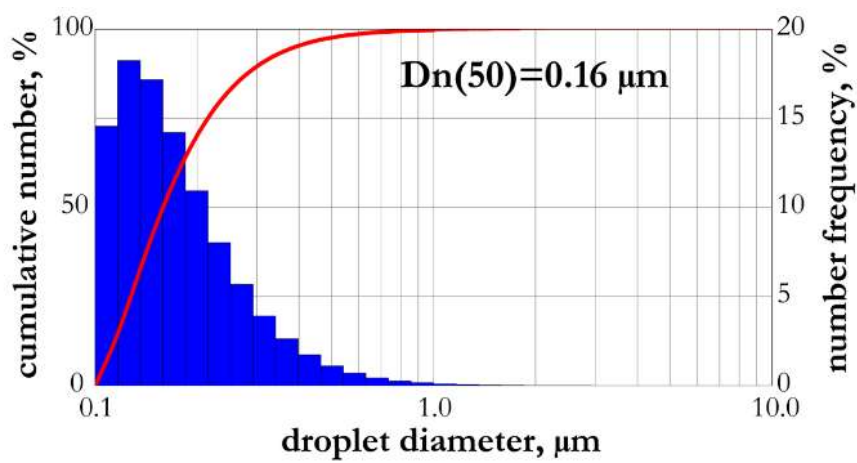

**Figure S1**. Typical number-weighted size distribution of water droplets obtained in liquid atomization process [1–4] generating submicron droplets by gas jet disintegration of thin liquid films formed as bubbles on a liquid surface.

# Thermodynamic analysis of the liquid atomization process.

The atomization process is analyzed by applying the law of mass conservation and the first law of thermodynamics for a control volume including the liquid and gas phases, as shown in Fig. 1b (see main paper text). The rate of change of mass, $m_{cv}$, in the control volume is determined by the difference of mass flow rates at the inlet, $\dot{m}_i$, and exit, $\dot{m}_e$:

$$\frac{dm_{cv}}{dt} = \dot{m}_i - \dot{m}_e. \tag{S.1}$$

The rate of change of the number of droplets, $n_{d,cv}$, in the control volume is given by the difference of the net rate of droplets entering at the inlet, $\dot{n}_{d,i}$, and exit, $\dot{n}_{d,e}$, plus the number rate of droplets produced in the control volume, $\dot{n}_d$:

$$\frac{dn_{d,cv}}{dt} = \dot{n}_{d,i} - \dot{n}_{d,e} + \dot{n}_d. \tag{S.2}$$

The rate of change of energy of the control volume, $E_{cv}$, is determined by the balance of the rate of (heat) energy supplied to the control volume, $\dot{Q}_{cv}$, the rate of work done by the control volume, $\dot{W}_{cv}$, and the difference of rates of enthalpy addition between the inlet, $\dot{H}_i$, and exit, $\dot{H}_e$ [5]:

$$\frac{dE_{cv}}{dt} = \dot{Q}_{cv} - \dot{W}_{cv} + \dot{H}_i - \dot{H}_e \tag{S.3}$$

The rate of work on the control volume is a sum of the rates of work done by the liquid, $\dot{W}_l$, and gas, $\dot{W}_g$, phases:

$$\dot{W}_{cv} = \dot{W}_l + \dot{W}_g \tag{S.4}$$

Assuming steady-state flow for a continuous adiabatic process at room temperature, neglecting the changes in potential energy of both fluids, evaporation and the change in liquid kinetic energy, assuming complete ideal expansion of the gas jet, disregarding drag and frictional energy dissipation, from the Eqs. (S.1), (S.2) and (S.3) we obtain $\dot{m}_i = \dot{m}_e$ and $\dot{W}_{cv} = 0$.

## Construction of atomization diagrams for droplet flow rates.

Atomization diagrams can also be constructed for droplet flow rates. For this purpose, we take time derivatives of all the terms of Eq. (1) given in the main paper text, and obtain that the balance of the rates of change of specific energy in the liquid atomization process shown in Fig. 1a is given by:

$$\dot{w}_{\Delta p} = \dot{w}_c + \dot{w}_v \tag{S.5}$$

Here $\dot{w}_{\Delta p}$ is the rate of specific work done by the atomizing gas on the liquid by means of the supplied differential gas pressure $\Delta p_{gj}$, and $\dot{w}_c$ and $\dot{w}_v$ are, respectively, the rates of specific work done by capillary and viscous forces in the atomized liquid. Employing the dimensionless analysis described in our previous article [2], we establish two dimensionless numbers $\mathrm{N}_{l,cv} = \rho_l^{5/2} \gamma_l^{3/2} d^{9/2} \xi / \mu_l^3$ and $\mathrm{N}_l = \Delta p_{gj}^{3/2} / \left( \rho_l \gamma_l^{3/2} d^{3/2} \xi \right)$, which determine the specific flow rate of droplets, $\xi = \dot{n}_{d,e} / \dot{m}_{d,e}$, for the expected droplet diameters $\mathrm{N}_d^{**} = \Delta p_{gj} \langle d \rangle / \gamma_l = 1$, and $\dot{n}_{d,e}$ and $\dot{m}_{d,e}$ are, respectively, the number and the mass flow rates of droplets at the outlet of the atomizer (see Fig. 1b). Performing the algebraic treatment with steps similar to those undertaken above for the droplet diameters, we conclude that in terms of the ratio of timescales of the involved phenomena, $\mathrm{N}_{l,cv} \sim \dfrac{\tau_v^3}{\tau_c^3}$ and $\mathrm{N}_l \sim \dfrac{\tau_c^3}{\tau_{\Delta p}^3}$. On the other hand, the scale of energy rate supplied by gas pressure is given by $\varepsilon_{\Delta p} = \ell^2 / \tau_{\Delta p}^3$, the scale of energy rate stored or released by surface tension is $\varepsilon_c = \ell^2 / \tau_c^3$, and the scale of energy rate dissipated by liquid viscosity $\varepsilon_v = \ell^2 / \tau_v^3$. We assume that ratios of energy rate scales are proportional to the corresponding ratios of rates of specific works, i.e., $N_{l,cv} \sim \dfrac{\varepsilon_c}{\varepsilon_v} = k_5 \dfrac{\dot{w}_c}{\dot{w}_v}$, and $N_l \sim \dfrac{\varepsilon_{\Delta p}}{\varepsilon_c} = k_6 \dfrac{\dot{w}_{\Delta p}}{\dot{w}_c}$, where $k_5$ and $k_6$ are coefficients of proportionality. Substituting these expressions into Eq. (S.5), we find:

$$N_l = k_6 + \frac{k_5 k_6}{N_{l,cv}} \qquad (S.6)$$

The established Eq. (S.6) relates the rates of specific work provided by the gas jets and dissipated by the atomized liquid in nondimensional form, and it determines the specific flow rates of the produced droplets. Again, as a first approximation, we assume that the proportionality coefficients are order of one, $O(k_5) = O(k_6) = 1$.

## Extended explanation for Figure 3a.

Figure 3a in the main paper text demonstrates the atomization diagram for flow rates of water droplets, which was constructed using Eq. (4) and by setting $k_6 = \pi/6$ and $k_5 k_6 = 1$. The former value was obtained by noticing the fact that in the limit $\mathrm{N}_{l,cv} \gg 1$, which holds for many regular liquids like water and aqueous solutions, we find from Eq. (4) that $\mathrm{N}_l^{**} = \Delta p_{gj}^{3/2} \big/ \left( \rho_l \gamma_l^{3/2} \langle d \rangle^{3/2} \langle \xi \rangle \right) = k_6$. By using $\langle \xi \rangle = \overline{\xi}_{1,0} = 1 \big/ \left( \dfrac{\pi}{6} \rho_l \overline{d}_{1,0}^3 \right)$, which is the natural connection between the expected specific flow rate, $\langle \xi \rangle$, and the expected droplet diameter, $\langle d \rangle$, given by $\mathrm{N}_d^{**} = \Delta p_{gj} \langle d \rangle \big/ \gamma_l = 1$, we obtain the equation for the expected specific droplet flow rate, $\mathrm{N}_l^{**} = \pi/6 = k_6$. Recalling that $\mathrm{N}_l \sim \dfrac{\varepsilon_{\Delta p}}{\varepsilon_c}$, we outline the atomization region of possible specific droplet flow rates by considering the expectation $\varepsilon_{\Delta p} \approx \varepsilon_c$, i.e., $O(\mathrm{N}_l) = 1$, and determine the boundaries by $\mathrm{N}_l = 0.1 \mathrm{N}_l^{**}$ (the lower boundary of insufficient atomization energy rate with $\varepsilon_{\Delta p} \ll \varepsilon_c$) and $\mathrm{N}_l = 10 \mathrm{N}_l^{**}$ (the upper boundary of excess of atomization energy with $\varepsilon_{\Delta p} \gg \varepsilon_c$). Using the experimental data [1], we construct Fig. 3a and observe good agreement between the theoretically expected and measured specific droplet flow rates; the experimental points either fall on or lie very close to the expectancy line, $\mathrm{N}_l^{**} = \pi/6$.

**Atomization diagrams for gasoline, diesel, and solutions of sodium alginate and sodium benzoate.**

In Figs. S2-S7 given below, the experimental points were obtained using the measured count mean droplet diameters, $\overline{d}_{1,0}$, at various differential pressures applied in the atomization process, $\Delta p_{gj}$, which were reported in Ref. [2,3]. Error bars denote the evaluated measurement uncertainties. In each diagram shown in Figs. S2-S6, three main regions are indicated by the respective boundaries and labels, as described in the main article text: (I) Atomization region, which is the region of expected and possible droplet diameters and specific droplet flow rates; (II) Insufficient atomization energy/energy rate, which is the region of impossible too small droplet diameters or too large specific droplet flow rates; (III) Excessive atomization energy/energy rate, which is the region of impossible too big droplet diameters or too small specific droplet flow rates. The Figs. S6 and S7 demonstrate the isolines corresponding to various values of the proportionality coefficients, $k_6$, in Eq. (4) of the main text.

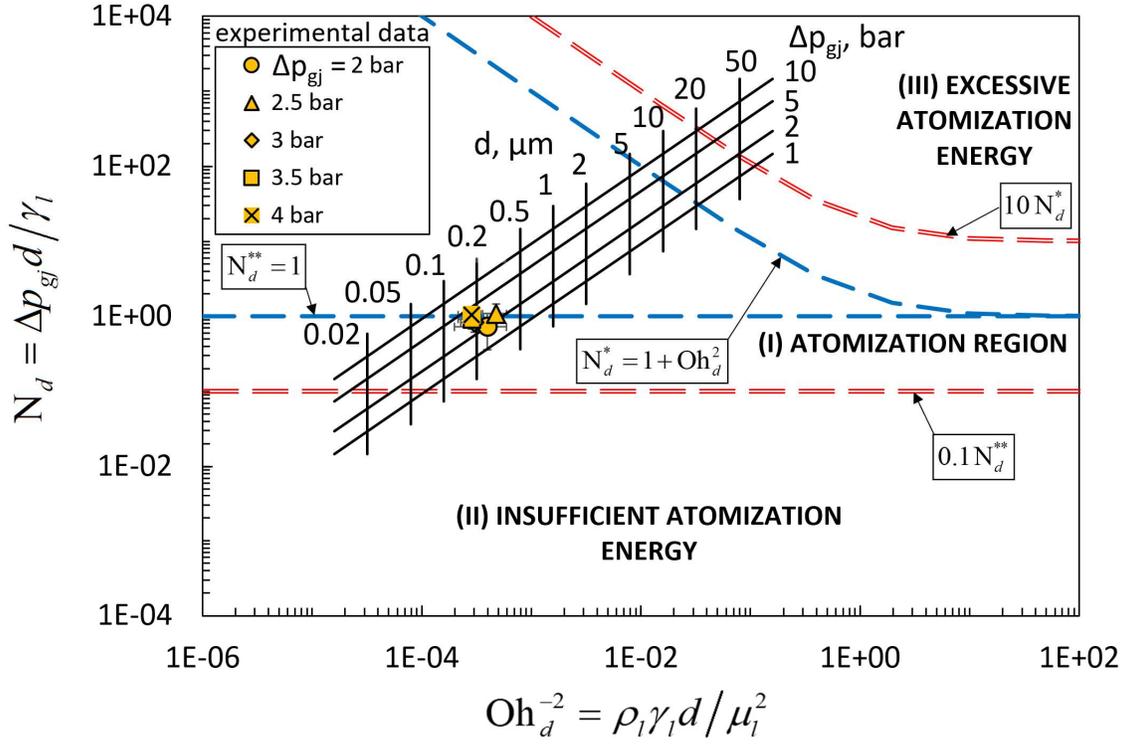

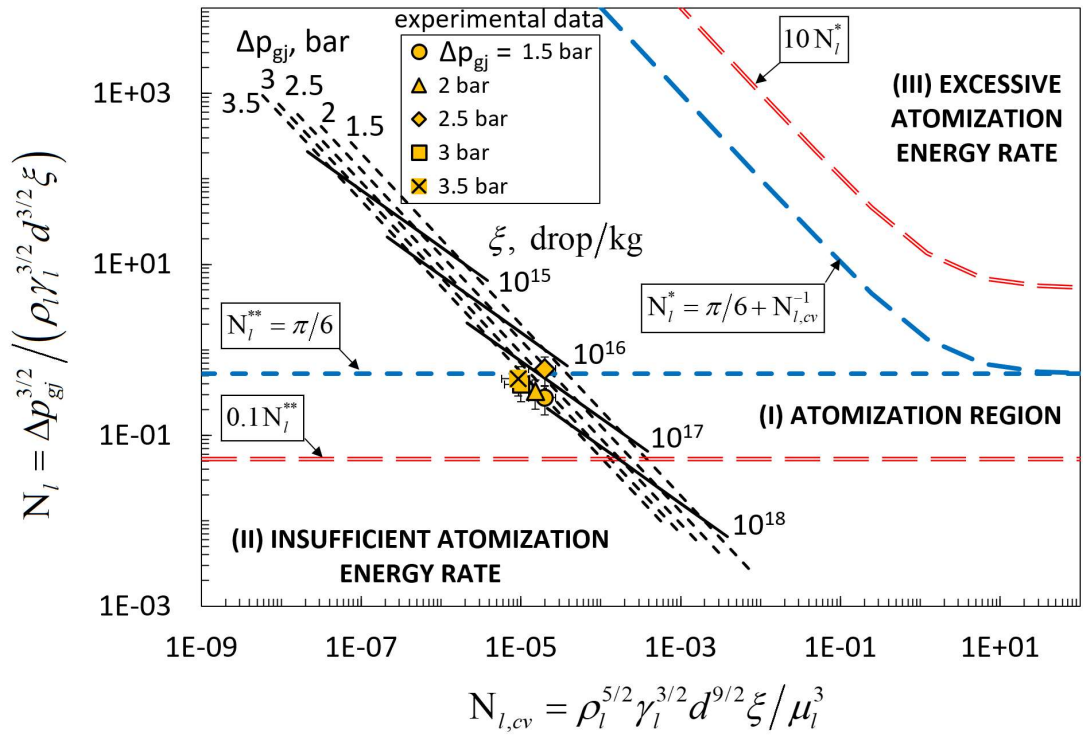

**Figure S2.** Atomization diagrams for droplets of aqueous solution of sodium alginate (2 wt%): (a) diameters, $d$, and (b) specific flow rates, $\xi$. The atomization process is schematically shown in Fig. 1a of the main text.

(a)

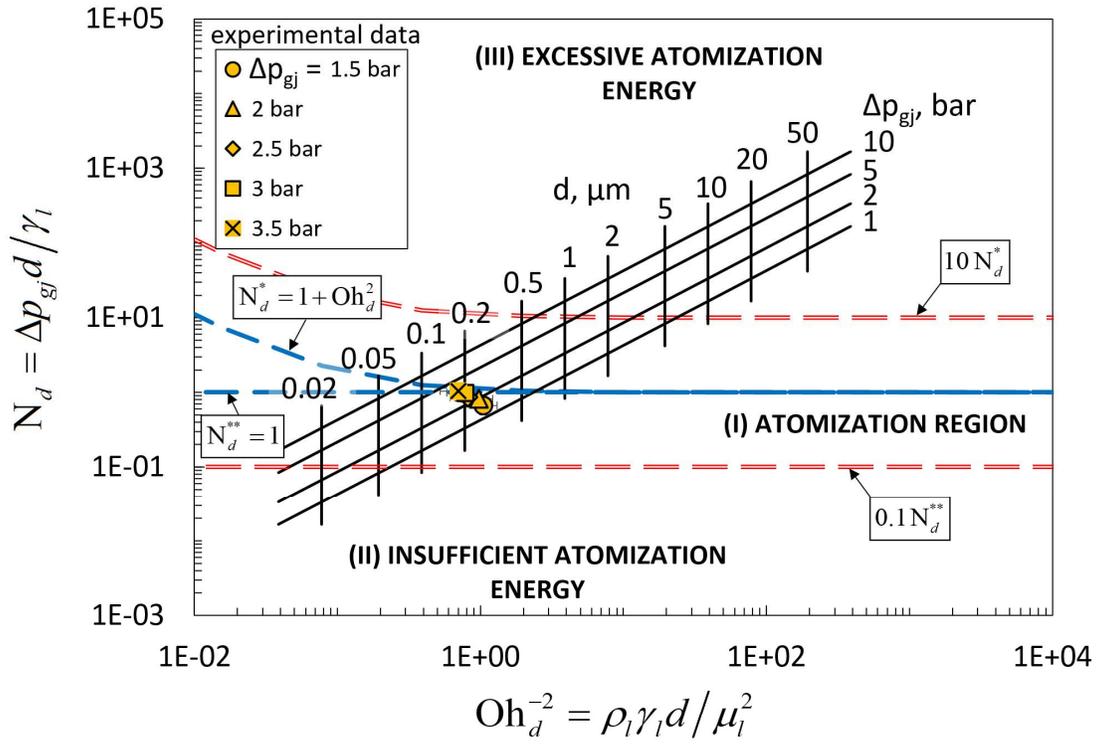

(b)

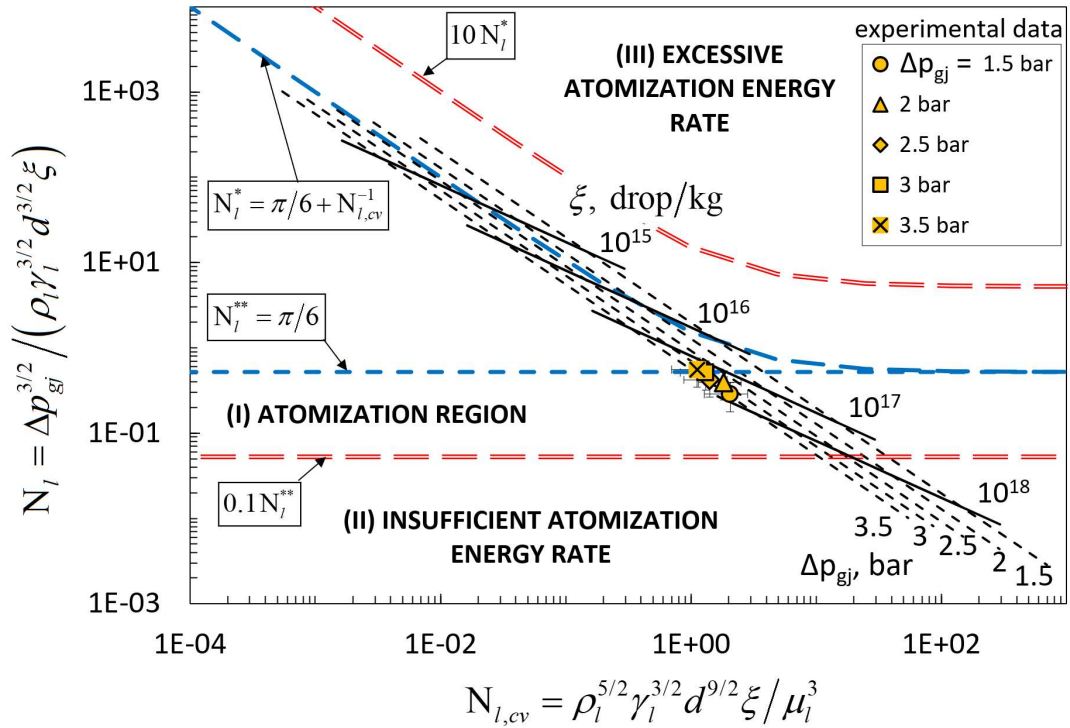

**Figure S3.** Atomization diagrams for droplets of aqueous solution of sodium benzoate (30 wt%): (a) diameters, $d$, and (b) specific flow rates, $\xi$. The atomization process is schematically shown in Fig. 1a of the main text.

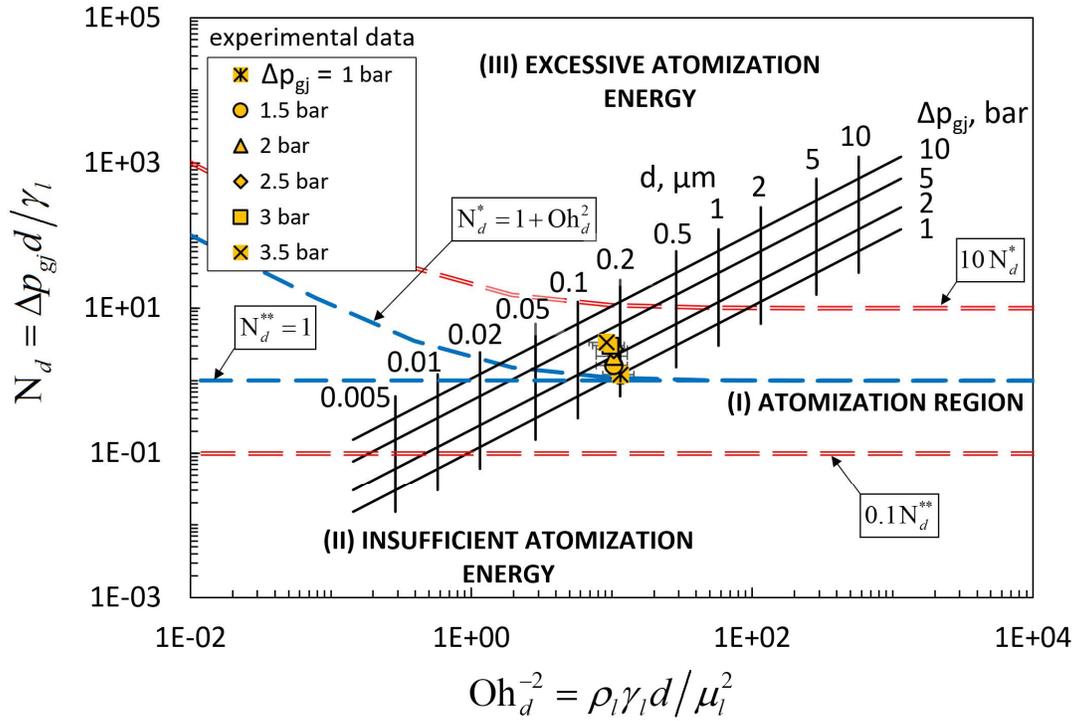

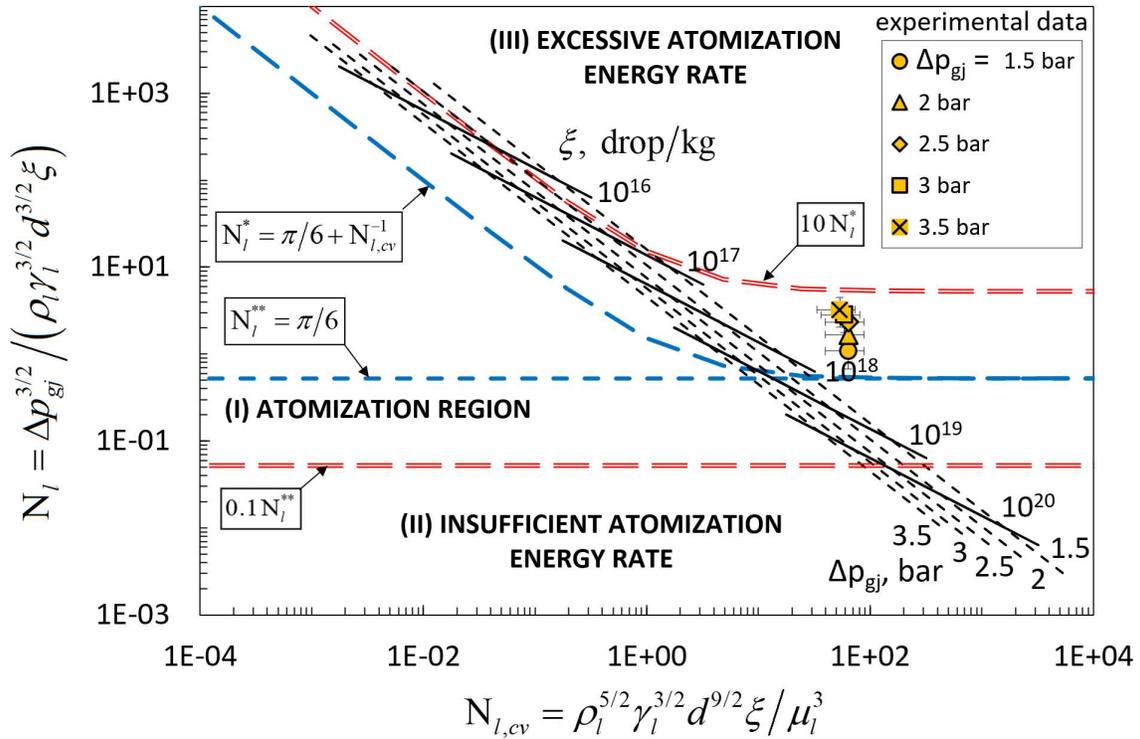

**Figure S4.** Atomization diagrams for gasoline droplets: (a) diameters, $d$, and (b) specific flow rates, $\xi$. The atomization process is schematically shown in Fig. 1a of the main text.

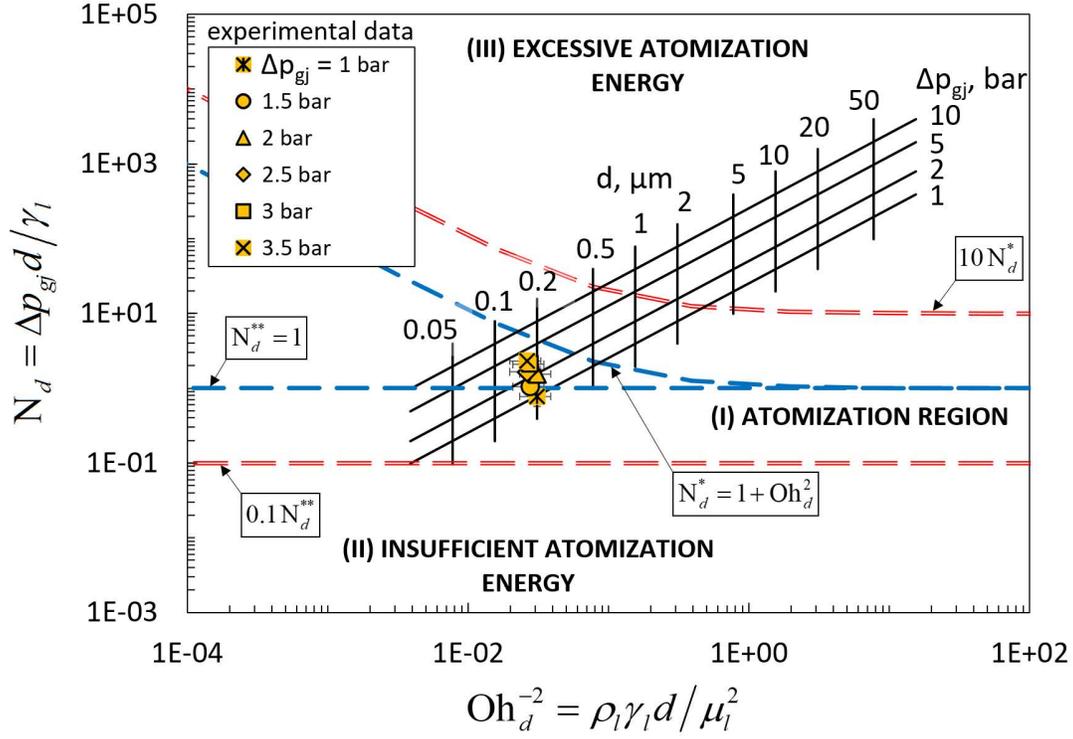

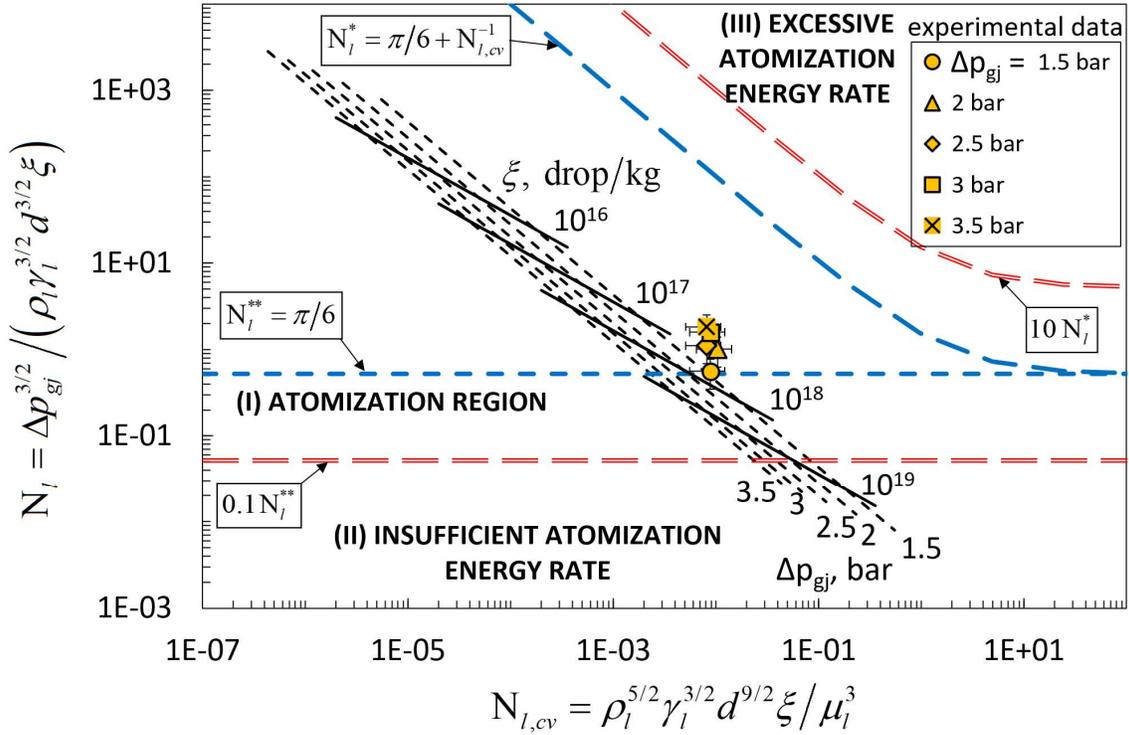

**Figure S5.** Atomization diagrams for diesel droplets: (a) diameters, $d$, and (b) specific flow rates, $\xi$. The atomization process is schematically shown in Fig. 1a of the main text.

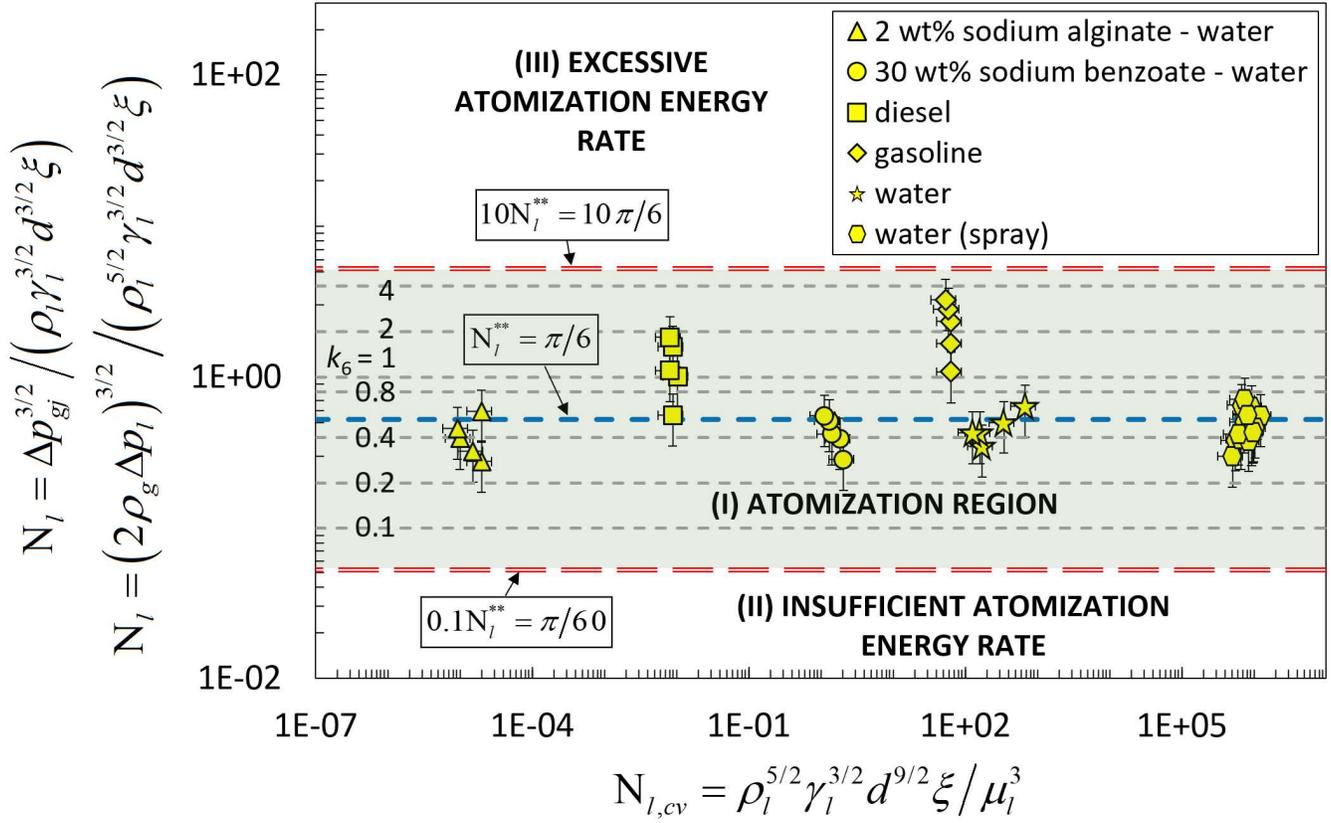

**Figure S6.** Atomization diagram summarizing specific droplet flow rates, $\xi$, produced by the two studied atomization processes: 1) aerosol generation shown in Fig. 1a of the main text, and 2) spraying from pressure nozzle given in Ref. [6].

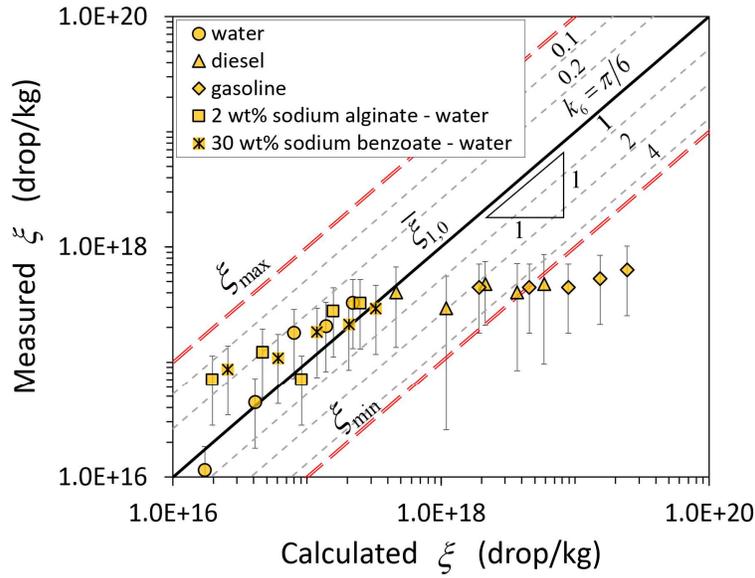

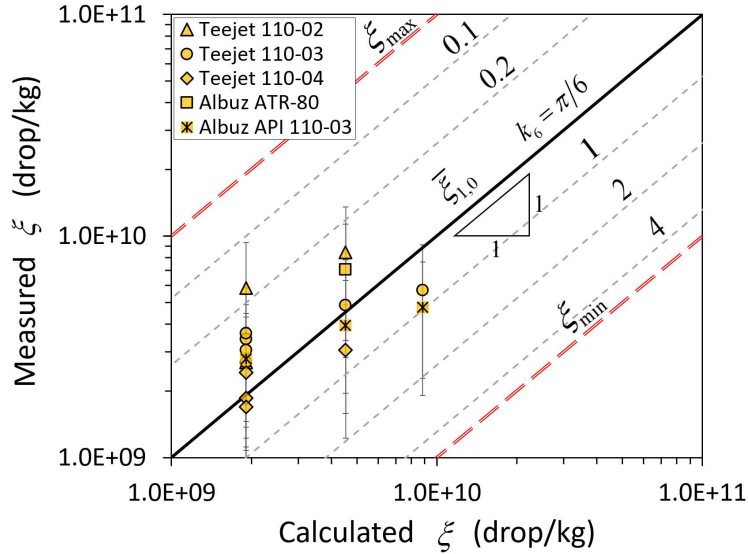

**Figure S7.** (a) Comparison between theoretical predictions and experimental data [1,2] for specific droplet flow rates, $\overline{\xi}_{1,0}$, of various liquids subjected to the atomization process shown in Fig. 1a of the main text. (b) Comparison between theoretical predictions and experimental data [6] for specific droplet flow rates, $\overline{\xi}_{1,0}$, of water subjected to atomization in various pressure nozzles. The lines $\overline{\xi}_{1,0}$ correspond to the lines of the expected dimensionless numbers $N_l^{**}$ in the atomization diagrams in Figs. S2b-S5b and Fig. S6, and of Fig. 3 of the main article. The double-dashed lines denote the theoretical lower and upper boundaries of possible specific flow rates, $\xi_{\min}$ and $\xi_{\max}$, for the produced polydisperse aerosols and sprays, and correspond to the boundary lines of the atomization regions in Fig. 3 of the main text. The diagonal isolines correspond to different values of the proportionality coefficient $k_6$ (see Eq. (S.6) Eq. (4) of the main text). Error bars indicate the evaluated measurement uncertainties.